\documentclass[]{interact}

\usepackage[caption=false]{subfig}
\usepackage{color}

\usepackage[numbers,sort&compress,merge]{natbib}
\bibpunct[, ]{[}{]}{,}{n}{,}{,}
\renewcommand\bibfont{\fontsize{10}{12}\selectfont}

\theoremstyle{plain}

\theoremstyle{definition}

\theoremstyle{remark}

\begin{document}

\articletype{ARTICLE TEMPLATE}

\title{Depletion-driven four-phase coexistences in discotic systems}

\author{
\name{
\'{A}lvaro Gonz\'{a}lez Garc\'{i}a\textsuperscript{a,b},
Remco Tuinier\textsuperscript{a,b},
Jasper V. Maring\textsuperscript{a,b},
Joeri Opdam\textsuperscript{a,b},
Henricus H. Wensink\textsuperscript{c},
and
Henk N.W. Lekkerkerker\textsuperscript{b}
}\affil{
\textsuperscript{a}Laboratory of Physical Chemistry, Department of Chemical Engineering and Chemistry, \& Institute for Complex Molecular Systems (ICMS)
Eindhoven University of Technology, P.O. Box $513$, $5600$ MB, Eindhoven, The Netherlands}\affil{
\textsuperscript{b}Van `t Hoff Laboratory for Physical and Colloid Chemistry, Department of Chemistry \& Debye Institute, Utrecht University, Padualaan $8$, $3584$ CH, The Netherlands}\affil{
\textsuperscript{c}Laboratoire de Physique des Solides - UMR $8502$, Universit\'{e} Paris-Sud, Universit\'{e} Paris-Saclay  and CNRS, $91405$ Orsay Cedex, France}
\thanks{CONTACT R.~T. Author. Email: r.tuinier@tue.nl}
}

\maketitle

\begin{abstract}
Free volume theory (FVT) is a versatile and tractable framework to predict the phase behaviour of mixtures of platelets and non-adsorbing polymer chains in a common solvent. Within FVT, three principal reference phases for the hard platelets are considered: isotropic (I), nematic (N), and columnar (C). We derive analytical expressions that enable us to systematically trace the different types of phase coexistences revealed upon adding depletants and confirm the predictive power of FVT by testing the calculated diagrams against phase stability scenarios from computer simulation. A wide range of \textit{multi}-phase equilibria is revealed, involving two-phase isostructural transitions of all phase symmetries (INC) considered as well as the possible three-phase coexistences. Moreover, we identify the system parameters, relative disk shapes and colloid-polymer size ratios, at which four-phase equilibria are expected. These involve a remarkable coexistence of all three phase states commonly encountered in discotics including isostructural coexistences I$_1$-I$_2$-N-C, I-N$_1$-N$_2$-C, and I-N-C$_1$-C$_2$. 
\end{abstract}


\section{Introduction}
Professor Daan Frenkel has developed ground-breaking insights into theoretical chemistry and statistical mechanics, including its application to soft matter\cite{frenkel_soft_2002,frenkel_colloid_2014} and in particular the phase behaviour of anisotropic particle dispersions\cite{eppenga_monte_1984,frenkel_thermodynamic_1988,frenkel_invited_1989,veerman_phase_1992,Frenkel_ellipsoids_1984,Bolhuis_1997}. He is undoubtedly best known for his computer simulation studies\cite{FrenkelBook_1996} but also has demonstrated deep insights into the theoretical aspects of the various subfields in which he is active. Daan has been at the forefront in using computer simulations to better understand transitions of both rodlike and discotic particles and has made seminal contributions to the understanding of the effect of depletants on the phase behaviour in particle suspensions\cite{bates_phase_2000,Meijer1994}. We therefore think it is appropriate to devote some recent insights we obtained on the phase behaviour of disc-polymer mixtures to this issue.

Since the pioneering work of Onsager in the 1940s\cite{onsager_effects_1949}, approaches that aim at understanding the effect of the colloidal particle shape on the thermodynamics of colloidal systems have gained increasing attention\cite{dijkstra_phase_2013}. The solvent can be treated as background in which the colloidal particles are assumed to interact only via hard-core repulsion\cite{poon_colloids_2016}. Phase transitions present in such systems are `entropy-driven', as there is no notion of internal energy when only hard interactions are accounted for\cite{FRENKEL199926,dijkstra_entropy-driven_2014}. The classical (nowadays text-book) example of such entropy-driven process is the isotropic to nematic phase transition of rods\cite{Bolhuis_1997,wensink_isotropicnematic_2003,roij_isotropic_2005}. For platelets, Frenkel himself has pioneered such entropy-driven phase transitions\cite{eppenga_monte_1984,frenkel_invited_1989,veerman_phase_1992,bates_infinitely_1998}, which receives increasing attention\cite{kouwer_specific_2003,harnau_inhomogeneous_2007,cuetos_columnar_2008,marechal_phase_2011,cienega-cacerez_induced_2016,tasios_simulation_2017}.

A next step to bring theoretical models towards real systems, which often comprise multiple components, is to consider a binary mixture of particles. It has been well-established that an effective depletion attraction takes place in asymmetric mixtures of particles (even for hard-sphere mixtures\cite{dijkstra_phase_1999}). Due to an imbalanced osmotic pressure that the smaller particles exert onto the bigger ones when the larger ones are close enough, an effective (indirect) attraction between the bigger components arises in a system where all direct interactions are repulsive\cite{van_anders_understanding_2014}. The first theory for the so-called depletion attraction originates from the work of Asakura and Oosawa in the  1950s\cite{Asakura1954,Asakura1958}. For colloidal spheres mixed with non-adsorbing polymer chains, this depletion attraction has been widely studied\cite{lekkerkerker_colloids_2011}, also for systems with direct colloid-colloid interactions beyond hard-core\cite{fortini_phase_2005,cinacchi_large_2007,garcia_tuning_2016}. 

Colloidal suspensions and colloid-polymer mixtures in which the colloidal particle has a plate-like shape are common in nature and technology: examples are pigments\cite{yuan_effects_2015}, clays\cite{bailey_smectite_2015}, blood\cite{ye_particle-based_2016},  food-stuffs\cite{Dickinson2016}, and disc-like polymer micelles\cite{lagerwall_nanotube_2007}. It has been experimentally observed how addition of non-adsorbing polymers\cite{van_der_kooij_phase_2000,Zhu_2007,Luan_2009} or small spheres\cite{oversteegen_crystallization_2005,kleshchanok_structures_2010,kleshchanok_attractive_2011,Doshi_2011,Kleshchanok_2012,de_las_heras_floating_2012,landman_effects_2014,chen_observation_2015} to a colloidal platelet suspension enriches the phase behaviour. For a review, see \cite{liu_graphene_2017}.

Theoretical models and computer simulations have predicted the experimentally observed Isotropic-Isotropic (I$_1$-I$_2$) and Nematic-Nematic (N$_1$-N$_2$) phase coexistences for platelets mixed with polymers or with small spheres. For infinitely thin plates, Bates and Frenkel\cite{bates_phase_2000} found I$_1$-I$_2$ and N$_1$-N$_2$ isostructural coexistences. Such I$_1$-I$_2$ and N$_1$-N$_2$ isostructural phase coexistences are a consequence of the depletion-induced attraction. A similar approach was followed by Zhang \textit{et al}. accounting for finite platelet thickness, and incorporating the columnar phase\cite{zhang_phase_2002,zhang_phase_2002-1}. Alternative approaches regarding platelet-sphere mixtures have also reported such isostructural coexistences\cite{harnau_structure_2008,de_las_heras_bulk_2013}. Aliabadi \textit{et al.}\cite{aliabadi_tracking_2016} presented a stability overview for hard platelets plates in a sea of hard spheres. Recently, we have reported a quadruple I$_1$-I$_2$-N-C phase coexistence for platelet-polymer mixtures\cite{Gonzalez2017} due to the merging of the I$_1$-I$_2$-N and the I$_1$-I$_2$-C triple coexistences. Such mechanism for quadruple coexistence had been recently exposed by  Akahane \textit{et al.}\cite{akahane_possible_2016} for single-component systems interacting via an anisotropic potential with variable tetrahedrality.

In this paper we show how the intricate effect of the excluded volume interactions leads to a rich phase behaviour for platelet-polymer mixtures. The theoretical approach developed recovers the results of previously-reported computer simulations and theories. On top of that, new coexisting phases are reported, namely a columnar-columnar (C$_1$-C$_2$) coexistence, and several types of three phase and four-phase coexistence regions, the latter being I$_1$-I$_2$-N-C, I-N$_1$-N$_2$-C, I-N-C$_1$-C$_2$. The structure of this paper is as follows. First, we report the basis of the theory developed, which provides the required tools to calculate the phase diagrams. Some results from our model are subsequently compared  to those emerging from other more numerically involved methods and with simulation approaches. We summarize our findings by presenting the various multiple phase equilibria that may occur in discotic colloid-polymer mixtures in a single, comprehensive plot spanned by the two relevant system parameters of our model, namely the disc aspect ratio and the colloid-to-polymer size ratio. Finally, we formulate the main conclusions.

\section{Theory and model comparison}
In this section the Free Volume Theory (FVT) developed to study the colloidal platelet-polymer mixtures of interest is briefly introduced. Firstly, we define the different length scales involved in our theoretical approach and recapitulate the known phase diagram for pure platelet suspensions. Subsequently, we briefly show how a semi-grand canonical approach enables to compute the phase behaviour of platelet-polymer mixtures. For the sake of completeness, technical details of the theory are presented in the Appendices. 

\subsection{Basics of the model}
We model the colloidal platelets as discs with diameter $D$ and thickness $L$.  The volume of the colloidal platelet ($v_c$) is given by $v_\text{c} = \frac{\pi}{4}D^2 L$. The aspect ratio $\Lambda = L/D$ defines the shape of the colloidal particle, and we focus on $\Lambda < 1$. These colloidal platelets are described as hard particles: overlap leads to an infinite repulsive interaction while the interaction is zero otherwise.

\begin{figure}[htb!]
    \centering
    \includegraphics[width=0.65\textwidth]{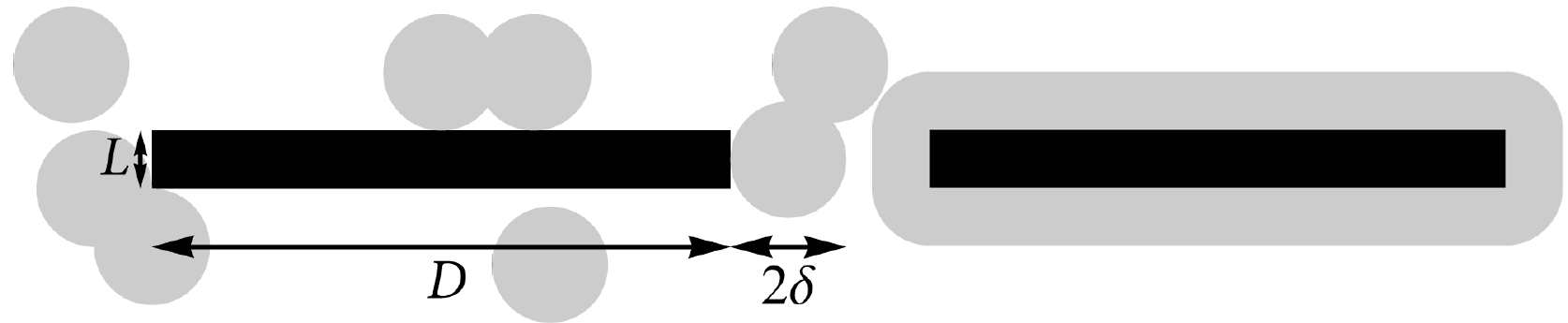} 
    \caption{Left: side view of a colloidal platelet with aspect ratio $\Lambda=L/D$ and a few PHS depletants with relative size $q = 2\delta/D$. The depletant radius ($\delta$), the platelet diameter ($D$), and the platelet thickness ($L$) are indicated. Right: corresponding depletion zone.}
\label{fig:ColDepl}
\end{figure}

The polymeric depletants are simplified as penetrable hard spheres (PHS) with radius $\delta$ (hence, their volume is $v_\text{d} = \frac{4\pi}{3}\delta^3$, and $\delta$  corresponds to the depletion thickness\cite{vrij_polymers_1976}). The PHS concentration is expressed via the dimensionless concentration, 
\begin{equation*}
\phi_\text{d}^\text{R} = \rho_\text{d}^\text{R} v_\text{d} \quad\text{,}
\end{equation*}
with $\rho_\text{d}^\text{R}$ the number density of depletants in bulk. These PHSs can penetrate each other, but are hard for the colloidal discs. The PHS model is a good approximation for polymers at low concentration and in a $\theta$--solvent\cite{tuinier_concentration_2004-1}. The relative size of the depletant with respect to the colloidal platelet is defined here as:
\begin{align*}
q = \frac{2\delta}{D} \quad\text{.}
\end{align*}
The three length scales introduced are schematized in Fig. \ref{fig:ColDepl}. The depletion zone surrounding any convex hard particle is enveloped by the surface with constant distance $\delta$ from the colloidal particle surface. Equivalently, the depletion zone is bounded by the Connolly surface\cite{connolly_molecular_1993} to the hard colloidal particle of interest. Consequently, the volume of the depletion zone ($v_\text{exc}$) corresponds to the excluded volume between a disc and a PHS\cite{oversteegen_general_2005}:
\begin{align}
v_\text{exc} &= 
\frac{\pi}{2}  D^2 \delta+
\frac{\pi}{4}  L (D+2\delta )^2+
 \frac{\pi^2}{2}  \delta ^2 \left(D+\frac{8\delta}{3\pi}\right)
\quad \text{.}
\label{eq:PHSPlateExc}
\end{align}
The length scales of interest and the typical shape of the depletion zone (side view) are depicted in Fig. \ref{fig:ColDepl}.

We present our expressions for the free energy ($\widetilde{F}_k$, with subscript $k$ running over the I, N and C states) of a system containing $N$ hard discs in a volume $V$ in terms of the volume fraction of platelets ($\phi_\text{c}$), and use dimensionless units:
\begin {align*}
\phi_\text{c} = \frac{N v_\text{c}}{V} \quad \text{;} \quad \widetilde{F}_k =\frac{  \beta F_k  v_\text{c}}{V} \quad \text{,}
\end{align*}
with $\beta = 1/(k_\text{B}T)$ the thermal energy in terms of the Boltzmann constant ($k_\text{B}$) and the absolute temperature $T$. Using the free energies for the different phase states\cite{wensink_phase_2009}, standard thermodynamic relations can be applied to calculate the osmotic pressure ($\Pi_k^\text{o}$) and chemical potential  ($\mu_k^\text{o}$)  of the pure platelet suspension in a given phase $k$; details are provided in Appendix \ref{ApPlatesExp}. 

\begin{figure}[htb!] 
    \centering
    \includegraphics[height=7.5cm]{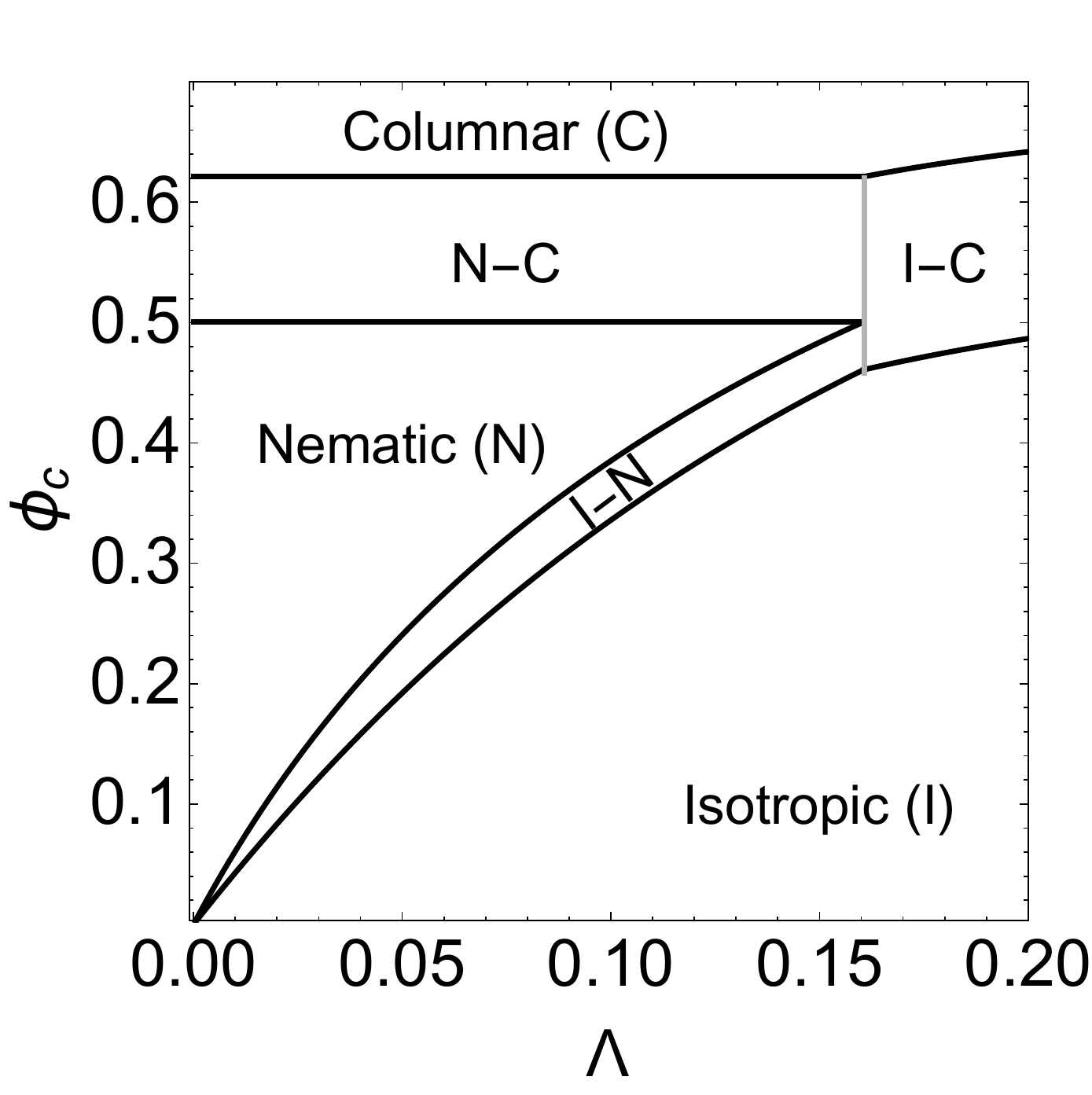} 
    \caption{Phase diagram of a monodisperse hard disc suspension in the $\{\Lambda\,\phi_\text{c}\}$ phase space. The long, grey triple line indicates the platelet aspect ratio beyond which Isotropic-Columnar (I-C) coexistence dominates over the I-N and Nematic-Columnar (N-C) using the Gaussian approximation for the phase ($\Lambda\approx 0.16$). At this specific $\Lambda$, there is an I-N-C triple point. A comparison of this phase diagram with computer simulation results (by Frenkel, among others) can be found in our previous work \cite{wensink_phase_2009}, and for more recent simulations see Marechal \textit{et al.} \cite{marechal_phase_2011}.
}
\label{fig:phasediagplates}
\end{figure}

This enables resolving the phase diagram for a system of hard platelets, presented in Fig. \ref{fig:phasediagplates}. The relatively high excluded volume between thin platelets explains the I-N phase transition occurring at very low packing fractions for very small values of the aspect ratio ($\Lambda \rightarrow 0$). With increasing $\Lambda$, the I-N coexistence widens and its boundaries shift towards higher packing fractions.  From Fig. \ref{fig:phasediagplates} it also follows that the N-C coexistence barely depends on $\Lambda$. For sufficiently thick discs ($\Lambda \approx 0.16$), transitions from an isotropic to a columnar phase occur without an intermediate nematic phase: thick discs are not sufficiently anisotropic to stabilize the occurrence of a nematic phase \cite{wensink_phase_2009}. The gray vertical line in Fig. \ref{fig:phasediagplates} at $\Lambda \approx 0.16$ indicates the I-N-C triple coexistence for hard colloidal platelets. The phase diagram presented in Fig. \ref{fig:phasediagplates} constitutes the reference point for understanding the thermodynamics of platelet-polymer mixtures. 

\subsection{Thermodynamics of platelet-polymer mixtures}
Free Volume Theory (FVT) is applied to compute the thermodynamic properties of colloidal platelets in a sea of polymer chains modelled as penetrable hard spheres (PHS) in a common solvent. We consider the platelet-polymer mixture of interest (the system, $\text{S}$) at a fixed temperature to be in osmotic equilibrium with a reservoir ($\text{R}$) where only depletants and solvent are present. We assume $\text{S}$ and $\text{R}$ to be separated  by a membrane permeable for the solvent and  depletants, but impermeable for the colloidal platelets. This defines a semi-grand potential $\Omega$, in which the (imposed) depletant concentration in $\text{R}$ ($\phi_\text{d}^\text{R}=\rho_\text{d}^\text{R}v_\text{d}$) fixes the chemical potential of the depletants in $\text{S}$, with $\rho_\text{d}^\text{R}$ the number density of depletants in $\text{R}$, since $\mu_\text{d}^\text{R} = \mu_\text{d}^\text{S}$. Note that fixing the chemical potential or the osmotic pressure in $\text{R}$ is equivalent to imposing a certain value for $\phi_\text{d}^\text{R}$, as no colloids are present in $\text{R}$. As we use the PHS approximation for polymers, we focus our attention on cases where $\phi_\text{d}^\text{R} < 1$, which actually covers the most interesting phase transitions found. 

We use dimensionless units for the grand-canonical ensemble:
\begin{align*}
\widetilde{\Omega}_k = \frac{\beta\Omega_k v_\text{c}}{V} \quad \text{.}
\end{align*}
The general expression for the semi-grand potential for ideal depletants \cite{lekkerkerker_phase_1992,lekkerkerker_colloids_2011} has the form:
\begin{align}
\widetilde{\Omega}_k(\phi_\text{c},\Lambda,q,\phi_d^\text{R}) &= \widetilde{F}_k(\phi_\text{c} ,\Lambda )- 
    \frac{v_\text{c}}{v_\text{d}}\alpha_k(\phi_\text{c} ,\Lambda ,q )\widetilde{\Pi}_\text{d}^\text{R}(\phi_d^\text{R})
    \quad \text{,}
\label{eq:Omega}
\end{align}
where $\alpha_k(\phi_\text{c} ,\Lambda ,q )$ is the free volume fraction for depletants in the system, and 
\begin{align*}
\widetilde{\Pi}_\text{d}^\text{R}(\phi_\text{d}^\text{R}) = {\beta \Pi_\text{d}^\text{R}v_\text{d}} \quad\text{,}
\end{align*}
is  the dimensionless osmotic pressure of an ideal solution of depletants in R, provided by  Van't Hoff's law: 
\begin{align*}
\widetilde{\Pi}_\text{d}^\text{R}(\phi_\text{d}^\text{R}) &=\phi_\text{d}^\text{R} \quad \text{.}
\end{align*}
The depletant volume fraction in $\text{S}$ follows from:
\begin{align*}
\phi_\text{d}^\text{S}&=\alpha_k(\phi_\text{c},\Lambda,q)\phi_\text{d}^\text{R} \quad \text{.}
\end{align*}
The free volume fraction for depletants in the system, $\alpha$, can be derived by combining  Widom's insertion method \cite{Widom1963} with Scaled Particle Theory (SPT) \cite{helfand_scaled_1960,lebowitz_scaled_1965} for the work required to bring a depletant from the reservoir to the system ($W$). Following the standard FVT approach (details of the calculation can be found in Appendix \ref{Ap:alfa}) we find:
\begin{align}
\alpha_k(\phi_\text{c},\Lambda,q) &= 
    (1-\phi_\text{c} )\exp \left[-Q_\text{s}(\phi_\text{c} ,\Lambda,q )\right]
    \exp \left[-\frac{v_\text{d}}{v_\text{c}}\widetilde{\Pi}_k^\text{o}(\phi_\text{c} ,\Lambda ) \right] \quad\text{,}
\label{eq:alfa}
\end{align}
with
\begin{align*}
Q_\text{s}(\phi_\text{c} ,\Lambda,q ) &= q \left(\frac{1}{\Lambda }+\frac{\pi  q}{2 \Lambda }+q+2\right) y(\phi_\text{c}) \\ &
+ 2 q^2 \left(\frac{1}{4 \Lambda ^2} +\frac{1}{\Lambda }+1\right) y(\phi_\text{c})^2 \quad \text{,}
\end{align*}
and 
\begin{align*}
y(\phi_\text{c}) = \frac{\phi_\text{c}}{1-\phi_\text{c}} \quad ,
\end{align*}
and where $\widetilde{\Pi}_k^\text{o}(\phi_\text{c} ,\Lambda )$ accounts for the depletant-free pressure of the phase state $k$ considered. Further details on the calculation of the phase diagrams are provided in Appendix \ref{diagcalc}.

\subsection{Model comparison}
\begin{figure}[htb!]
    \centering
    \includegraphics[height=4.5cm]{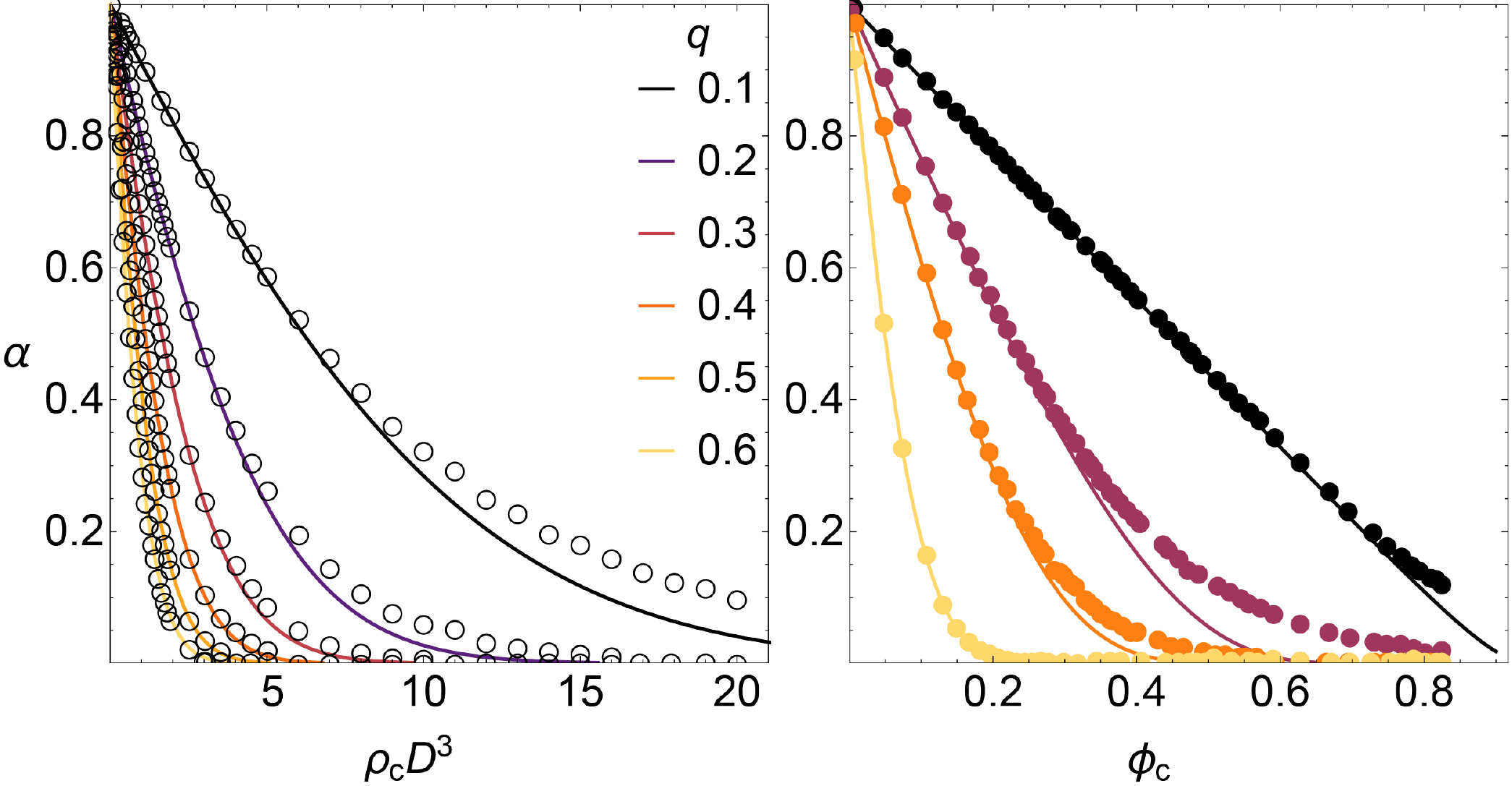}
    \caption{Comparison of the free volume fraction using Eq. \ref{eq:alfa} using the osmotic pressure of an isotropic platelet phase (curves) with available computer simulations literature data (symbols). Left: infinitely thin platelets ( simulation data from\cite{bates_phase_2000}) compared with $\Lambda = 10^{-5}$ in our approach. Right: cut-spheres with $\Lambda = 0.1$ (simulation data from\cite{zhang_phase_2002-1}). The relative size of the depletant $q$ is indicated in the inset. Note that on the left panel we keep the units as in the original reference.}
\label{fig:alfa2}
\end{figure}

We first compare our model both at the free volume level and at the phase diagram level with available computational approaches. In Fig. \ref{fig:alfa2} the calculated free volume fractions (using the osmotic pressure of the isotropic phase) from Eq. \ref{eq:alfa} are compared with previous results from computer simulations\cite{bates_phase_2000,zhang_phase_2002,zhang_phase_2002-1}. Our results are in concordance with simulation data both for infinitely thin plates\cite{bates_phase_2000} and for finite size cut-spheres\cite{zhang_phase_2002,zhang_phase_2002-1} mixed with PHS. The qualitative agreement between our theoretical approach and previous simulations supports the accuracy of the derived expression for the free volume fraction of PHSs in a suspension of hard discs.

Phase diagrams for infinitely thin plates mixed with PHS are compared with those calculated by Bates and Frenkel\cite{bates_phase_2000}. We use dimensionless number density of platelets along the abscissa ($\rho_\text{c}D^3 = 4{\phi_\text{c}}/(\pi\Lambda)$) and the fugacity of depletants ($z=\rho_\text{d}^\text{R}D^3=\phi_\text{d}^\text{R}q^{-3}6/\pi$) for the ordinate for comparison purposes. In their approach, the free volume fraction for PHSs in a disc suspension is fitted from simulation results, and FVT is applied for the calculation of the phase diagrams considering the full numerical solution of the disk orientation probability in the nematic phase. As the difference in the I-N coexistence between Odijk's Gaussian trial function approximation and the self-consistent numerical approach for the nematic phase of infinitely thin plates is quite pronounced, the depletant-free baselines ($z=0$) deviate from each other. However, the phase sequences occurring at each range of depletion attraction are not affected by the choice of the orientational distribution probability for the nematic phase.

\begin{figure*}[htb!]
    \centering
    \includegraphics[height=9cm]{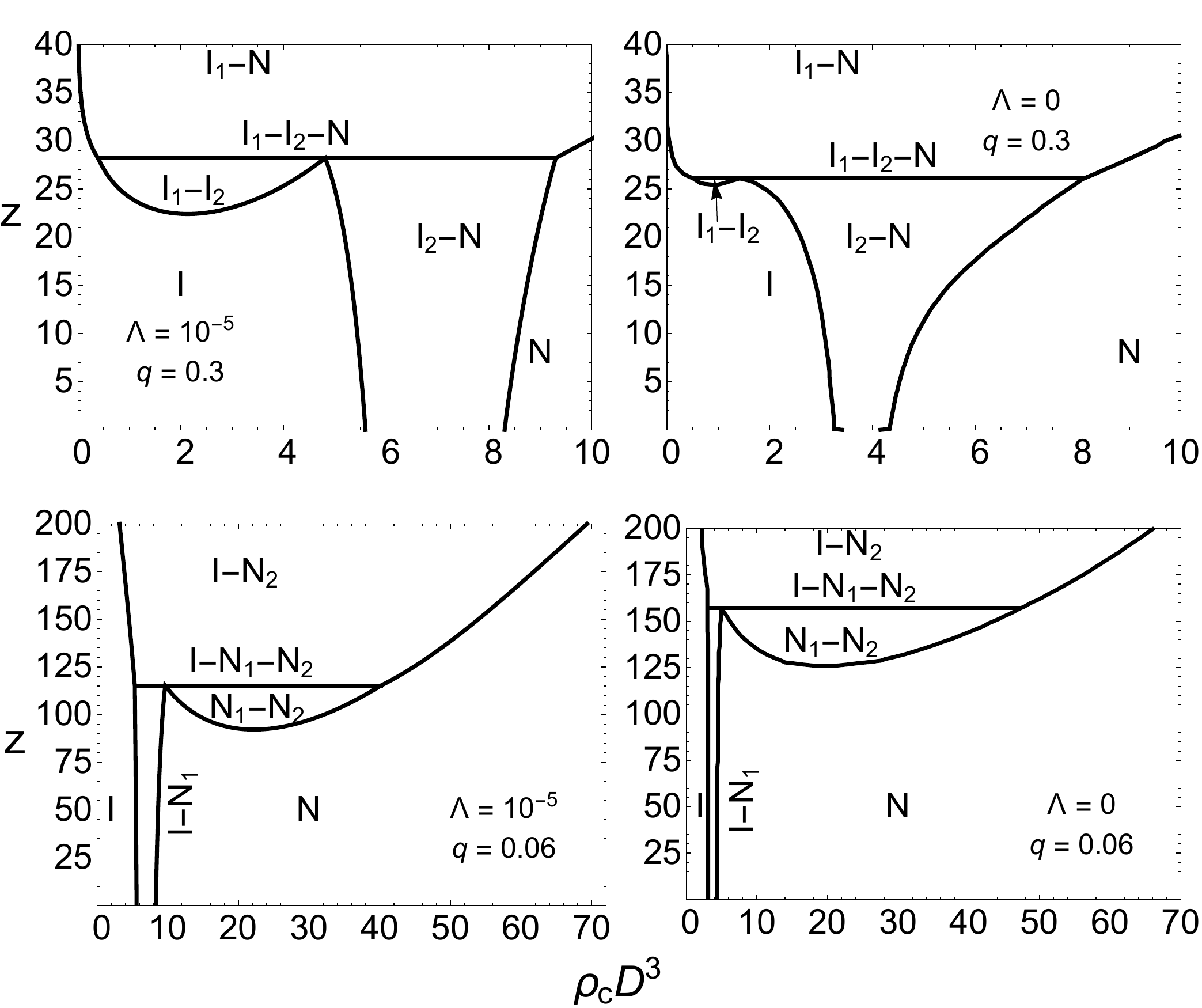}
    \caption{Phase diagrams from FVT (left panels) compared to results by Bates and Frenkel\cite{bates_phase_2000} (right panels).}
\label{fig:Comp2000}
\end{figure*}

\begin{figure}[htb!]
    \centering
    \includegraphics[height=9cm]{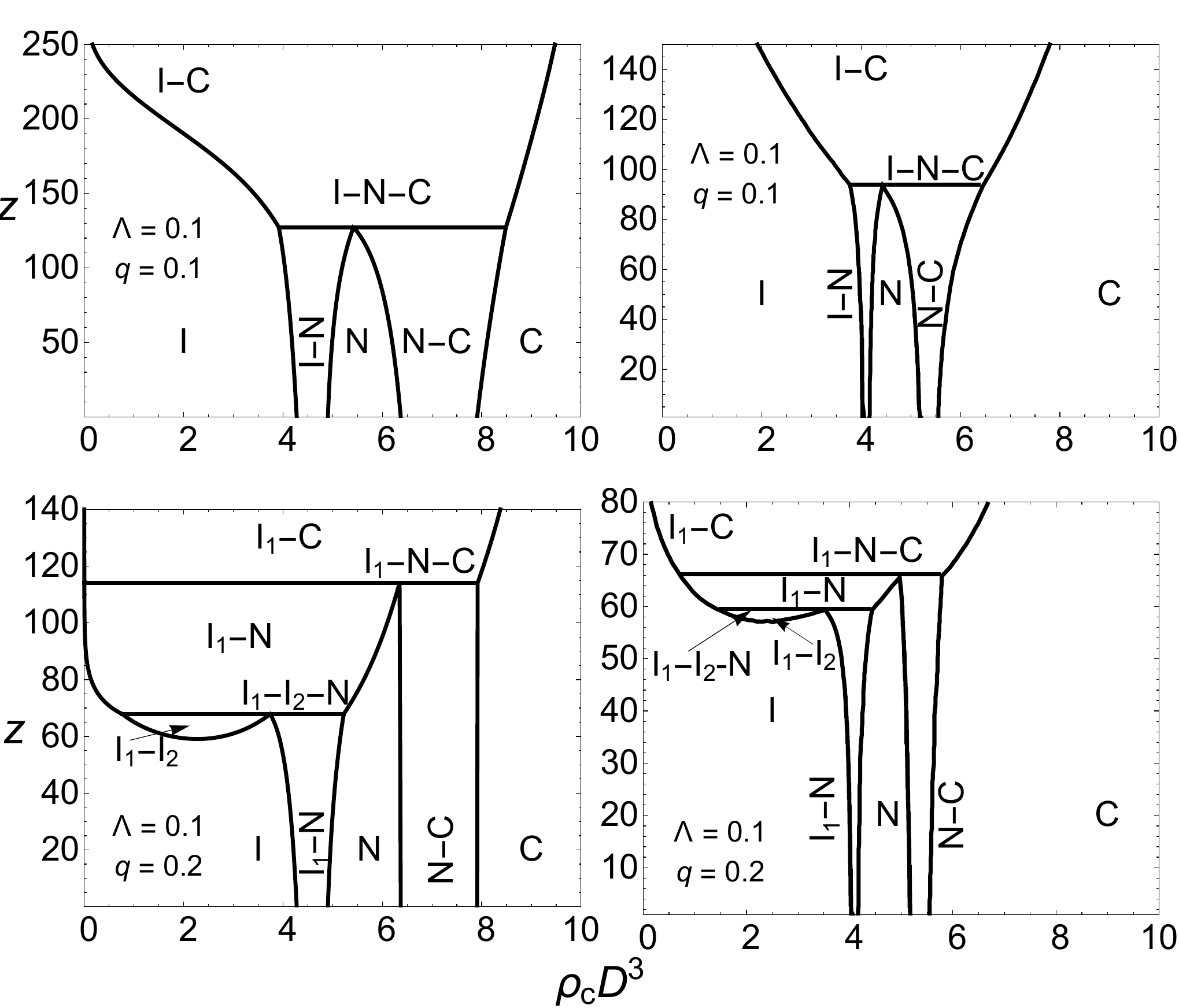}
    \caption{Phase diagrams obtained from our FVT results for hard discs + PHS (left panels) compared to the phase diagrams for cut spheres + PHS from Zhang \textit{et al.}\cite{zhang_phase_2002} (right panels).}
    \label{fig:Comp2002}
\end{figure}

Next, we compare our theoretical approach with the phase diagrams generated by mixtures of thick platelets (described as cut-spheres) plus depletants by Zhang \textit{et al}.\cite{zhang_phase_2002} for $\Lambda=0.1$ and $q=\{0.1,0.2\}$ (see Fig. \ref{fig:Comp2002}). The hybrid Monte Carlo-FVT approach followed by Zhang \textit{et al}.\cite{zhang_phase_2002} is similar to the one of Bates and Frenkel\cite{bates_phase_2000}. Even though the free volume fractions ($\alpha$, see Fig. \ref{fig:alfa2}) are quite close, the depletant fugacity that leads to phase coexistence is higher for our colloidal discs as for these cut-spheres. This is due to the higher excluded volume between discs as between cut-spheres. Note again that the overall topology of the phase diagram from FVT agrees with the one obtained from the hybrid method even though the platelet mesogen differs. 

\section{Multi-phase coexistences and critical end points}
In a large part of the $\{\Lambda,q\}$ parameter space we find phase diagrams such as these on the top panels of Fig. \ref{fig:Comp2002}. Isostructural phase transitions (I$_1$-I$_2$, N$_1$-N$_2$, and C$_1$-C$_2$) lead to a considerable enrichment of the phase diagrams including six three-phase coexistences: I$_1$-I$_2$-N, I$_1$-I$_2$-C, I-N$_1$-N$_2$,N$_1$-N$_2$-C, I-C$_1$-C$_2$, and N-C$_1$-C$_2$; and three four-phase regions: I-N$_1$-N$_2$-C, I-N$_1$-N$_2$-C, and I-N-C$_1$-C$_2$. In this section we delineate the regions in the $\{\Lambda,q\}$ plane where three- and four-phase regions occur and discuss why these appear. To gain insight into the different types of multi-phase coexistence regions involving isostructural coexistence it is useful to focus on the isostructural phase coexistences in the vicinity of their corresponding critical endpoints\cite{LANG1975190,Warren1997,garcia_tuning_2016}. The critical endpoint (CEP) is defined as the condition under which the critical point of the isostructural phases is in equilibrium with a distinct third phase\cite{Widom_1973}.

\subsection{Isostructural isotropic coexistence}\label{sec:II}
We first consider the I$_1$-I$_2$ coexistence and check under which conditions it coexists with other phases. The $q$ value at which the critical point of the I$_1$-I$_2$ coexistence meets the I$_1$-I$_2$-N three phase region marks the relative range of the depletion interaction below which I$_1$-I$_2$ transition becomes metastable with respect to the I-N transition. Above this $q$ value we have a stable three-phase I$_1$-I$_2$-N region. Similarly,  the $q$ value at which the critical point of the  I$_1$-I$_2$ region meets the I$_1$-I$_2$-C three phase region marks the relative range of the depletion interaction below which I$_1$-I$_2$ transition becomes metastable with respect to the I-C transition; above this $q$ value three-phase I$_1$-I$_2$-C equilibria appear. Due to the decrease of the platelet-platelet excluded volume with increasing $\Lambda$, the range of the depletion attraction required for an I$_1$-I$_2$-N multi-phase coexistence lowers as the discs become thicker. However, when the columnar state dominates over the nematic one at sufficiently high $\Lambda$ and $q$, the scenario changes and the I$_1$-I$_2$ coexistence is connected to a I$_1$-I$_2$-C triple point.

\begin{figure*}[htb!]
    \centering
    \includegraphics[width=.97\textwidth]{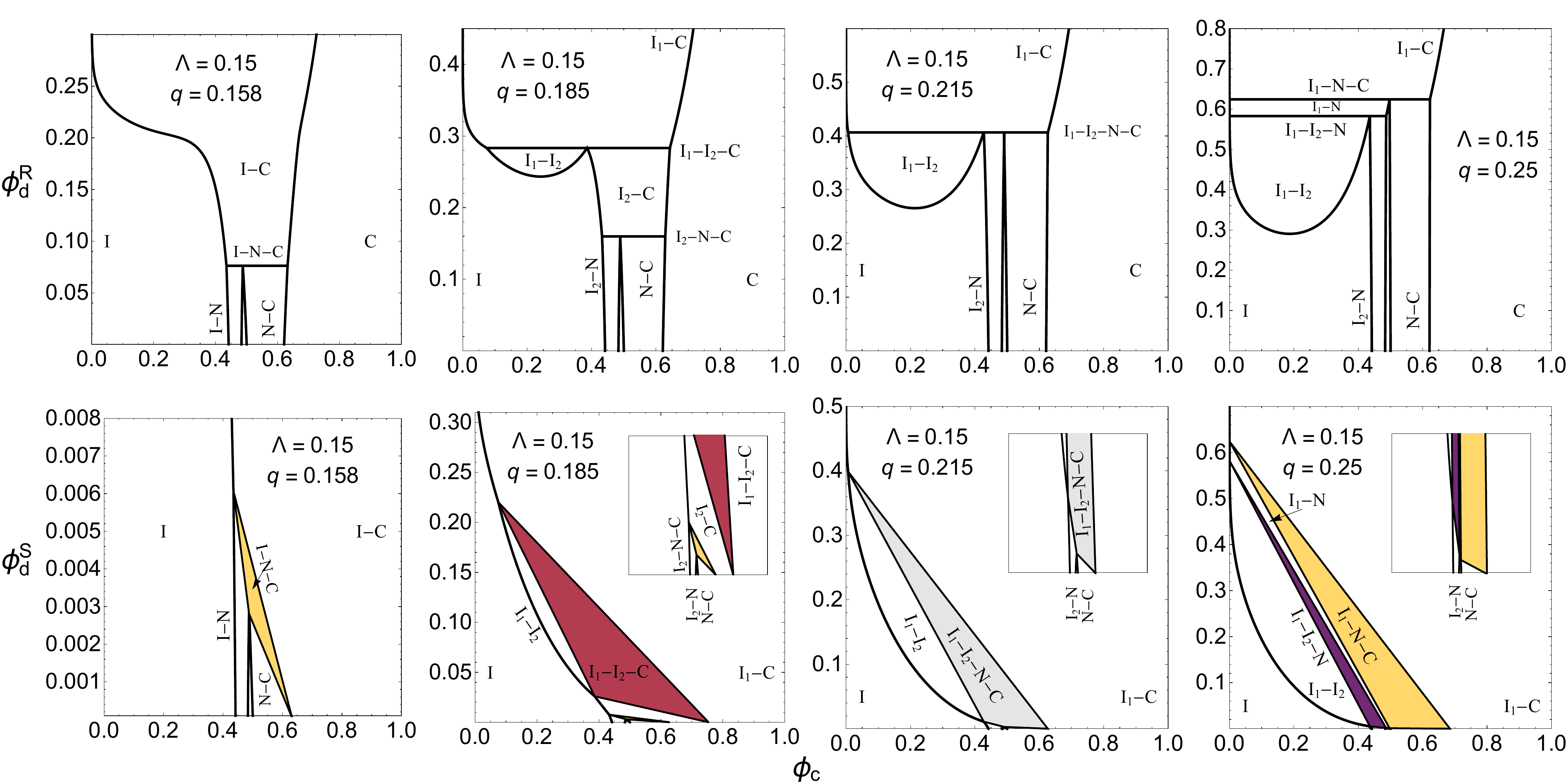}
    \caption{Collection of binodal phase diagrams for platelet-polymer mixtures in the $\{\phi_\text{c},\phi_\text{d}^\text{R}\}$ phase space for $\Lambda = 0.15$ and various $q$ values as indicated. Horizontal lines mark multiple-phase coexistence (more than two-phase). Bottom panels: as top ones but in the $\{\phi_\text{c},\phi_\text{d}^\text{S}\}$ phase space. The coloured triangles indicate the system representation of the triple-point lines on the top panels. Inset plots zoom into the low-depletant concentration regime.}
\label{fig:QPIINCPlot}
\end{figure*}

The calculated (I$_1$I$_2$)-N and (I$_1$I$_2$)-C CEP's (shown in Fig. \ref{fig:CEP_MasterPlot}) coincide at $\{\Lambda,q\}\approx\{0.122,0.163\}$, leading to a remarkable (I$_1$I$_2$)-N-C CEP where the critical endpoint of the I$_1$-I$_2$ critical point is in equilibrium with \textit{two} distinct phases: N and C. For $\Lambda = 0.15$ the transitions from no I$_1$-I$_2$ ($q = 0.158$) to I$_1$-I$_2$-C ($q = 0.185$) to I$_1$-I$_2$-N ($q = 0.25$) can be observed in Fig. \ref{fig:QPIINCPlot} (upper panels), where the free volume theory (FVT) binodals are plotted for various $q$ values. A value of $q = 0.158$ does not provide an attraction long-ranged enough to induce an isostructural I$_1$-I$_2$ coexistence. For $q = 0.185$, stable I$_1$-I$_2$ coexistence is possible, and an I$_1$-I$_2$-C triple line can be observed above the I-N-C triple line, while for $q = 0.25$ an I$_1$-I$_2$-N triple line appears at lower $\phi_\text{d}^\text{R}$ than the I$_1$-N-C triple line. Strikingly, at $q = 0.215$ a I$_1$-I$_2$-N-C four phase coexistence is predicted: at this $q$ value the I$_1$-I$_2$-N, I$_1$-N-C, I$_1$-I$_2$-C and I$_2$-I-C three phase  lines merge; we reported this quadruple coexistence recently.\cite{Gonzalez2017}. This I$_1$-I$_2$-N-C four-phase coexistence occurs along the curve of $\{\Lambda,q\}$ values from the (I$_1$I$_2$)-N-C CEP($\{\Lambda,q\} \approx \{0.122, 0.163\}$) towards the triple-line for platelets in the absence of depletants ($\Lambda \approx 0.16$). In the system representation (bottom panels of Fig. \ref{fig:QPIINCPlot}), the area of the multi-phase coexistence denotes the region where they are predicted in terms of colloid and depletion concentrations for the particular set of relevant size ratios.

\subsection{Isostructural nematic coexistence}\label{sec:NN}
\begin{figure*}[htb!]
    \centering
    \includegraphics[width=.97\textwidth]{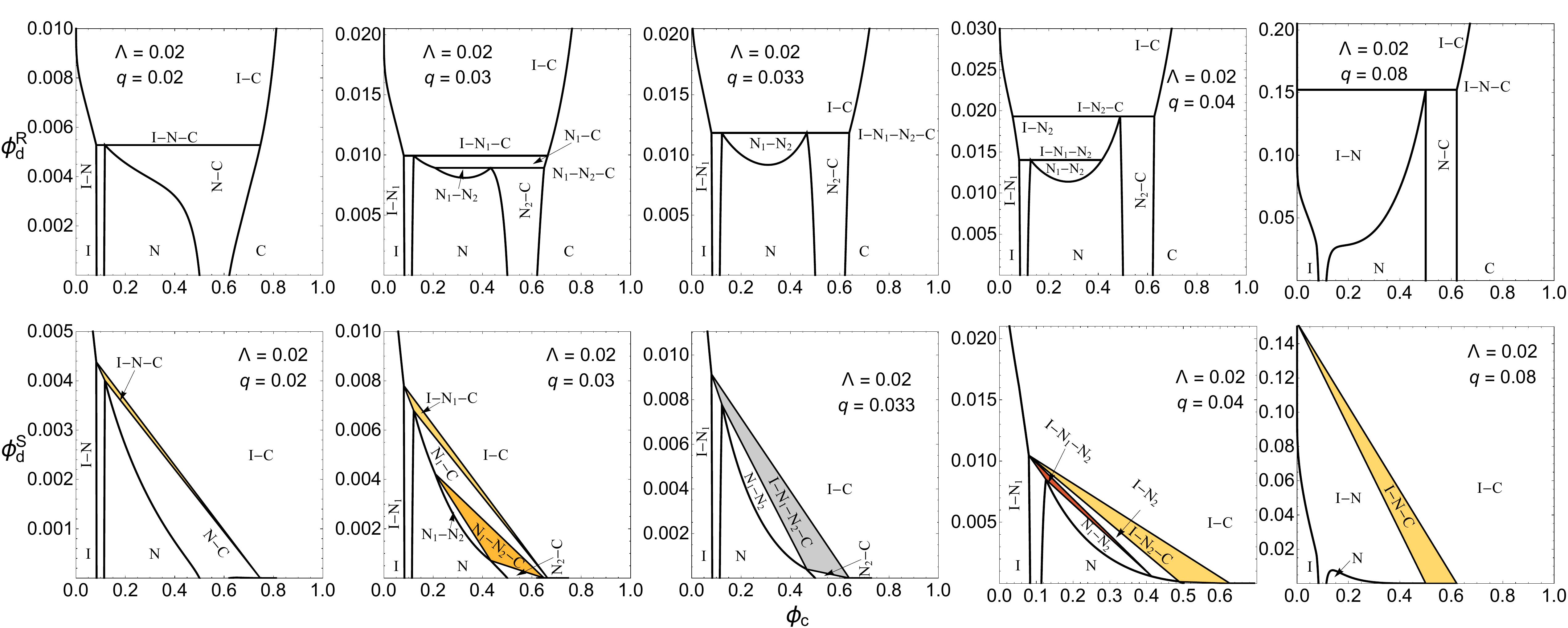}
    \caption{Same as Fig. \ref{fig:QPIINCPlot}, but for very thin hard plates ($\Lambda = 0.02$) and small polymers. An isotructural nematic-nematic coexistence occurs only within a certain range depletant sizes.}
\label{fig:QPINNCPlot}
\end{figure*}

We now pay attention to the N$_1$-N$_2$ coexistence and check under which conditions it coexists with other phases. 
Again we first consider the N$_1$-N$_2$ CEP's I-(N$_1$N$_2$) and (N$_1$N$_2$)-C. From previous calculations\cite{moradi_study_2017} it is known that N$_1$-N$_2$ equilibria are only stable at low values of $\Lambda$ and $q$. In Fig.\ref{fig:QPINNCPlot} we illustrate the influence of varying $q$ at $\Lambda=0.02$. It appears there is only a limited range of $q$-values where N$_1$-N$_2$ equilibria are found. At low $q$ values, the N$_1$-N$_2$ transition becomes metastable with respect to the N-C transition and at somewhat higher value of $\Lambda$ and increasing $q$ values the N$_1$-N$_2$ coexistence becomes metastable with respect to the I-N transition. For $q = 0.02$, there is no  N$_1$-N$_2$ coexistence, for $q = 0.03$ stable N$_1$-N$_2$ coexistence region is possible and a N$_1$-N$_2$-C triple line emerges, while for $q = 0.04$ a I-N$_1$-N$_2$  triple line is present. For $q=0.08$ there is again no N$_1$-N$_2$ coexistence.  For $\Lambda=0.2$ and $q=0.033$ the N$_1$-N$_2$-C, I-N$_1$-C, I-N$_1$-N$_2$ and I-N$_2$-C three-phase lines merge and again a four phase equilibrium is predicted: a quadruple I-N$_1$-N$_2$-C coexistence.  In the bottom panels of Fig. \ref{fig:QPINNCPlot}, the areas denote the regions in the system representation where multi-phase coexistences are predicted in terms of colloid and depletion concentrations for the particular set of relevant size ratios.

\subsection{Isostructural columnar coexistence}\label{sec:CC}
\begin{figure*}[htb!]
    \centering
    \includegraphics[width=.97\textwidth]{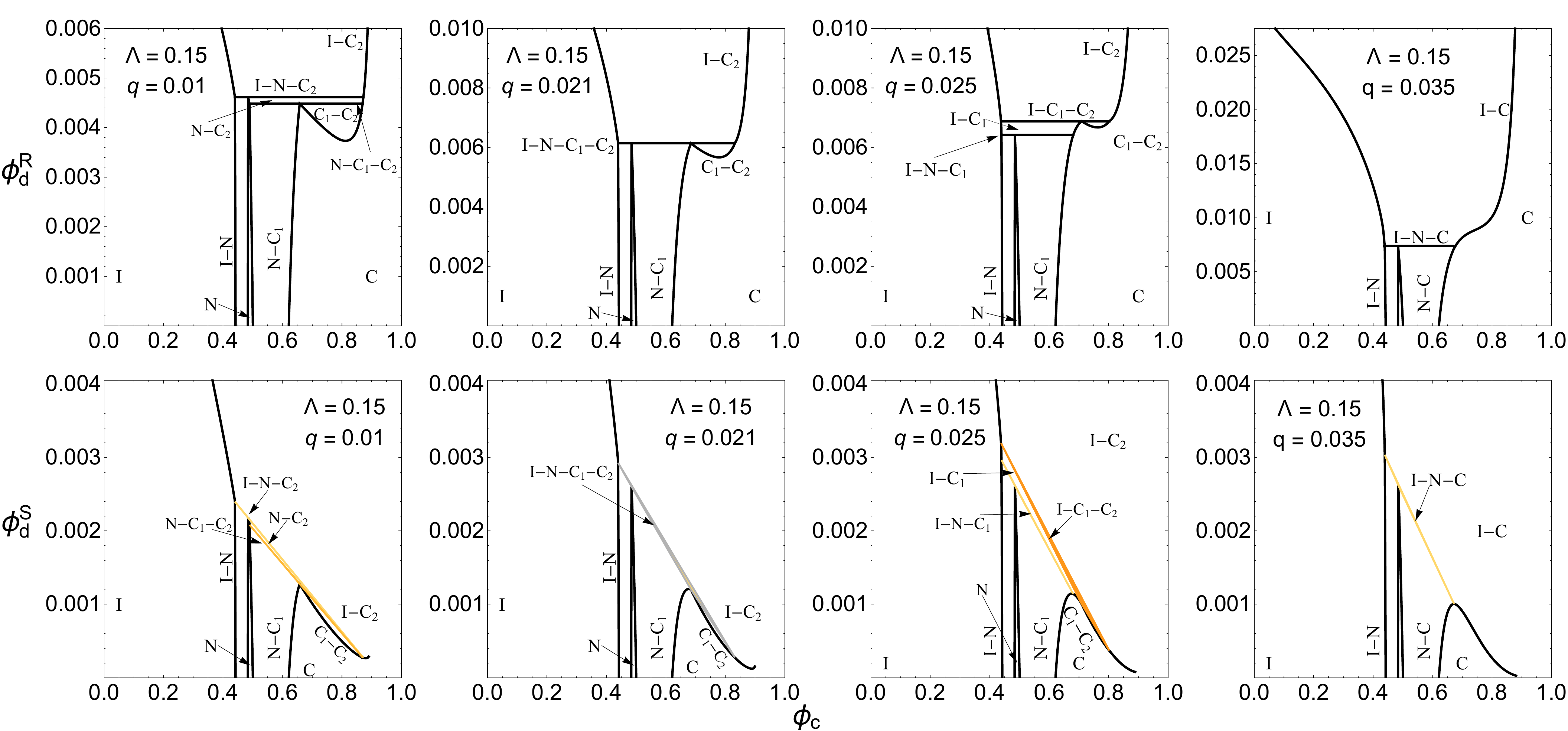}
    \caption{Same as Fig. \ref{fig:QPIINCPlot} for the $q$ indicated. For thick platelets and small polymers an isotructural columnar-columnar coexistence occurs.}
\label{fig:QPINCCPlot}
\end{figure*}

Surprisingly, systematically scanning the possible phase equilibria also revealed isostructural columnar phase state coexistence regions. Therefore the focus now is on C$_1$-C$_2$ coexistences, and we investigate under which conditions isostructural columnar equilibria coexists with other phase. In Fig. \ref{fig:QPINCCPlot} we present phase diagrams around for $\Lambda = 0.15$. For $q= 0.01$ a N-C$_1$-C$_2$ triple line emerges while for $q = 0.025$ an I-C$_1$-C$_2$ triple line is present. For $q=0.035$ there is no longer C$_1$-C$_2$ coexistence. For $q=0.021$ the N-C$_1$-C$_2$, I-N-C$_1$, I-C$_1$-C$_2$ and the I-N-C$_2$ three-phase lines merge and again a four-phase equilibrium is predicted: the I-N-C$_1$-C$_2$ coexistence. This four-phase coexistence occurs along a curve of $\{\Lambda,q\}$ values from the I-N-(C$_1$C$_2$) CEP at $\{\Lambda,q\}\approx\{0.135,0.029\}$ towards $\{\Lambda,q\}\approx\{0.155,0\}$. In the system representation (bottom panels of Fig. \ref{fig:QPINCCPlot}), the area of multi-phase coexistence denotes the region where they are predicted in terms of colloid and depletion concentrations for the particular set of relevant size ratios. {Isostructural} C$_1$-C$_2$ coexistence have {not} been reported yet, higher order multi-phase coexistences for this isostructural coexistence are unknown as well. 

\subsection{Multi-phase coexistence overview}\label{sec:masterplot}
\begin{figure*}[htb!]
    \centering
		\includegraphics[height=7.5cm]{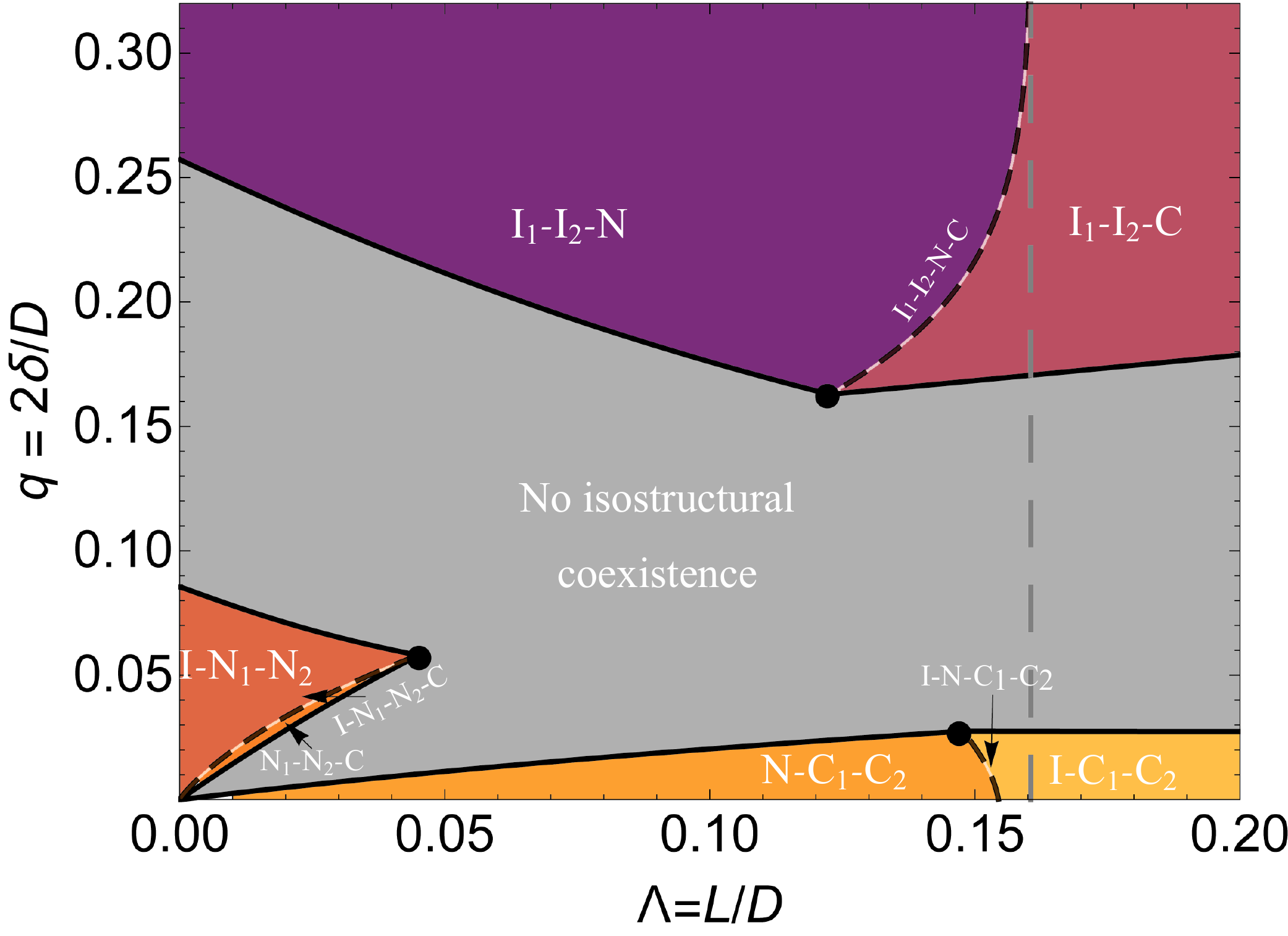}
    \caption{Generic overview of isostructural multi-phase coexistences found in a colloidal platelet-polymer mixture in terms of the platelet aspect ratio ($\Lambda$) and the relative size of depletant ($q$). Solid curves correspond to critical end point (CEP) curves of the isostructural phase coexistences as indicated. Black dots indicate the points where two critical endpoint (CEP) curves coincide, thus defining the CEP of the quadruple coexistences. Quadruple coexistences (dashed curves) bound the corresponding isostructural coexistence regions. Long-dashed, grey, vertical line corresponds to the triple point (I-N-C) for platelets in the absence of depletants ($\Lambda \approx 0.16$). Note that three-phase coexistence regions correspond to areas in the $\{\Lambda,q\}$ phase space, whereas four-phase coexistences are only found along their respective curves. The tiny region marked with an arrow corresponds to the N$_1$-N$_2$-C coexistence.} 
\label{fig:CEP_MasterPlot}
\end{figure*}

We calculated the possible thermodynamically stable phases via the critical end point (CEP) of the different isostructural critical points with the corresponding triple points. The results are summarized in Fig. \ref{fig:CEP_MasterPlot}. In the infinitely thin platelet limit, the stable phases found correspond to those of Bates and Frenkel\cite{bates_phase_2000}. The trends with increasing platelet thickness correspond to hybrid theory/simulation approaches for cut-spheres mixed with PHS\cite{zhang_phase_2002,zhang_phase_2002-1}. The area covered by the I-N$_1$-N$_2$ triple equilibrium matches with previous theoretical studies\cite{aliabadi_tracking_2016}, and reveals possible new regions for multi-phase coexistence. At $\{\Lambda, q\} \approx \{0.045, 0.058\}$, the I-(N$_1$N$_2$) and (N$_1$N$_2$)-C  CEP lines coincide, hence defining the I-(N$_1$N$_2$)-C CEP leading to a small nose-like region in which the N$_1$-N$_2$ transition is stable (see Fig. \ref{fig:CEP_MasterPlot}). This region is in agreement with the results reported by Aliabadi \textit{et al.}\cite{moradi_study_2017} using a canonical Onsager-Parsons-Lee approach for colloidal discs-hard sphere mixtures to obtain the boundaries of the stable isostructural N$_1$-N$_2$ transition regions. The I-N$_1$-N$_2$-C four-phase coexistence occurs along a curve of $\{\Lambda,q\}$ values from the I-(N$_1$N$_2$)-C CEP at $\{\Lambda, q\} \approx \{0.045, 0.058\}$ towards the corner of the $\{\Lambda,q\}$ diagram. This quadruple curve practically coincides with the (N$_1$N$_2$)-C CEP curve (see Fig. \ref{fig:CEP_MasterPlot}). Hence, the area covered by the N$_1$-N$_2$-C three-phase coexistence is very small.

The isostructural C$_1$-C$_2$ coexistence takes place at very low $q$ values (as shown in Fig.\ref{fig:CEP_MasterPlot}). This is reminiscent of the occurrence of two isostructural solid phases in a system of spheres with a very narrow range attraction\cite{bolhuis_isostructural_1994,dijkstra_phase_1999}. Again we first consider the C$_1$-C$_2$ CEP's: I-(C$_1$C$_2$) and N-(C$_1$C$_2$). As we saw in section \ref{sec:NN} with increasing $\Lambda$ and low $q$ values the N$_1$-N$_2$ transition becomes  metastable with respect to the N-C transition. On further  lowering $q$ we obtain the N-(C$_1$C$_2$) CEP (see Fig. \ref{fig:CEP_MasterPlot}). With further increasing $\Lambda$ the isotropic state takes over from the nematic in this competition, and we obtain the I-(C$_1$C$_2$) CEP (see Fig. \ref{fig:CEP_MasterPlot}). At $\Lambda=0.135$ and $q=0.25$ the I-(C$_1$-C$_2$) and  N-(C$_1$-C$_2$) CEP curves coincide, and we obtain  the I-N-(C$_1$C$_2$) CEP. The increasing in possible $q$ values leading to C$_1$-C$_2$ with $\Lambda$  can be further rationalized in terms of the depletion between the sides of the discs: while the appearance of columnar phases at $\Lambda$ values where they were not present in the pure platelet suspension is attributed to alignment of the flat phases leading to the maximum depletion attraction, within each hexagonal slice of the columnar phase the depletion attraction increases with increasing platelet thickness. This enables the columnar phase to separate into two phases for small enough depletants, promoting one phase with more free volume for depletants coexisting with a less depletant-concentrated but denser in platelets columnar state. The fact that there is rather limited space, even in the dilute columnar phase, explains that C$_1$-C$_2$ equilibria are only possible for relatively small depletant sizes (small $q$ values).

The overview shown in Fig. \ref{fig:CEP_MasterPlot} constitutes the main result of this research: it provides a systematic overview of where the possible critical points, three-, and four-phase coxistence areas occur in terms of the (geometrical) system parameters $\{\Lambda,q\}$.

It may be questioned whether four-phase coexistences violate the Gibbs phase rule, which dictates that in a system of $\mathfrak{C}$ components the maximum number of phases that can coexist is $\mathfrak{C}+2$. As shown by Tanaka and coworkers\cite{akahane_possible_2016}, for systems with an additional model parameter, four phases can coexist even in single component systems for a specific value of that parameter. In the system considered here there are only hard interactions between the two components (solvent is considered as background) so the temperature is effectively not a degree of freedom. Thus the maximum number of phases that can coexist is $\mathfrak{C}+1$. So at first sight with $\mathfrak{C}=2$ a four phase equilibrium does not seem possible. However, here we have two additional parameters at our disposal, $\Lambda$ and $q$, which act as additional field variables, thus allowing more than $\mathfrak{C}+1$ phases to be simultaneously in equilibrium, at appropriately constrained values of these parameters (see Fig.\ref{fig:CEP_MasterPlot}). {Fig. \ref{fig:CEP_MasterPlot} shows that for this particular system it is not possible to find coexistence between more than four phases. For other particle shapes (e.g. biaxial plates providing an additional system parameter) it might be possible to find a quintuple or even a sextuple point. A mixture showing two isostructural transitions (e.g. N$_1$-N$_2$ and C$_1$-C$_2$) simultaneously could perhaps lead to such conditions.}

In the experiments where colloidal platelets, which in the pure state give rise to liquid crystal phases, are mixed with nonadsorbing polymers\cite{van_der_kooij_phase_2000,Zhu_2007,Luan_2009} or with small spheres\cite{oversteegen_crystallization_2005,kleshchanok_structures_2010,kleshchanok_attractive_2011,Doshi_2011,Kleshchanok_2012,de_las_heras_floating_2012,landman_effects_2014,chen_observation_2015} the phase behaviour changes significantly compared to that of the system of pure platelets. We have to realise that the depletion interaction in combination with the inherent polydisperse nature of the colloids\cite{Bates_polydisperse_1999,van_der_kooij_phase_2000,Wensink_2004} and the sedimentation equilibrium under gravity\cite{Wensink_sediment_2004,Heras_stack_2013} may give rise to the presence of additional multi-phase coexistences. {For example Wensink and Lekkerkerker\cite{Wensink_sediment_2004} showed that a four phase equilibrium I$_1$-I$_2$-N-C may arise in the gravitational field even if the system without gravity only displays the two phase equilibria I$_1$-I$_2$, I$_2$-N, N-C. De las Heras \textit{et al.}\cite{de_las_heras_floating_2012,Heras_stack_2013} showed that in a system with only one isotropic phase without gravity in the gravitational field two isotropic layers may appear with a nematic phase floating in between}. Moreover, most experimental colloidal systems are non-hard and contain additional direct interactions. Here we briefly consider the experimentally observed multiphase coexistences for mixtures of colloidal platelets with nonadsorbing polymers.

Van der Kooij \textit{et al.}\cite{van_der_kooij_phase_2000} studied a system of sterically stabilised Gibbsite platelets dispersed in toluene mixed with the nonadsorbing polymer polydimethylsiloxane(PDMS). In this plate-polymer mixture $\Lambda\approx 0.07$ and $q\approx 0.35$ (values of $\Lambda$ and $q$ given here refer to the ratios of the average thickness and diameter of the platelets and the average diameters of the polymers and platelets respectively). The sterically stabilised Gibbsite platelets in their pure state exhibit a I-N transition and a N-C transition with increasing concentration\cite{kooij_liquid_2000}. In the Gibbsite-PDMS mixture a four-phase equilibrium I$_1$-I$_2$-N-C, bordered by three three-phase equilibria I$_1$-I$_2$-N,I$_1$-N-C, and I$_2$-N-C was observed. The results were rationalised by representing the polydisperse platelets by a bidisperse system consisting of platelets of lower $\Lambda$ and higher $\Lambda$. As we have seen in our calculations (see Fig. \ref{fig:CEP_MasterPlot}) this gives rise to the three-phase regions  I$_1$-I$_2$-N and I-N-C for the lower $\Lambda$ and the three-phase regions  I$_1$-I$_2$-C and I-N-C for the higher $\Lambda$. In the three-dimensional concentration diagram spanned by the concentrations of the lower $\Lambda$ platelets, the higher $\Lambda$ platelets and the polymer, the four three-phase regions intersect and a tetrahedron-shaped four-phase I$_1$-I$_2$-N-C region appears bordered by four three-phase regions: I$_1$-I$_2$-N, I$_1$-N-C, I$_1$-I$_2$-C and I$_2$-N-C. The projection of these four- and three-phase regions on the experimental plane then leads the observations referred to.
 
Zhu \textit{et al.}\cite{Zhu_2007} studied a system of positively charged Mg$_2$AI layered double hydroxide platelets mixed with polyethylene glycol (PEG). Those platelets exhibit an I-N transition\cite{Liu_2003}. In this plate-polymer mixture $\Lambda \approx 0.055$ and $q \approx 0.01$. A multiphase coexistence consisting of a dilute upper phase, two or three birefringent phases and amorphous bottom phase was observed. The results are ascribed to a combination of the depletion interaction and sedimentation equilibrium. The nature of the birefringent phases was not determined; however in one experiment a phase which gives Bragg reflections of visible light was observed indicating that there is positional order in this phase. So this phase may be a columnar phase,  which given the values of $\Lambda$ and $q$ in this experiment would agree with the results given in Fig. \ref{fig:CEP_MasterPlot}.

Luan \textit{et al.}\cite{Luan_2009} observed in a system of positively charged Mg$_2$AI layered double hydroxide platelets mixed with polyvinyl pyrrolidone (PVP) with $\Lambda \approx 0.026$ and $q \approx 0.78$ multiphase coexistences consisting of up to six phases: a dilute upper phase, two isotropic phases, two nematic phases and an amorphous bottom phase and two or three birefringent phases. They ascribe these results to a combination of the depletion interaction, sedimentation equilibria and polydispersity effects.

\section{Conclusions}
We demonstrate that free volume theory (FVT) constitutes an efficient and tractable thermodynamic framework capable of unravelling the complicated multiphase behaviour of disc-polymer mixtures. As a result, a multi-phase coexistence overview in terms of the platelet thickness ($\Lambda$) and the relative depletant size ($q$) was obtained (Fig. \ref{fig:CEP_MasterPlot}). The possible phase states of the canonical platelet system were considered: isotropic (I), nematic (N) and columnar (C). The final phase diagrams do not only match with previous theoretical approaches and with experimental results but also exhibit a columnar-columnar isostructural coexistence not reported before. On top of the recently reported I$_1$-I$_2$-N-C quadruple coexistence, two further four-phase coexistences are put forward involving orientationally ordered isostructural coexistences at low depletant sizes: I-N$_1$-N$_2$-C, and I-N-C$_1$-C$_2$. All quadruple coexistences arise when two different isostructural triple phase coexistences merge. The stability regions can be explained in terms of excluded volume repulsions between hard discs being reduced by the second component in the mixture. The appearance of columnar phases can be rationalized in terms of alignment (and stacking) of the flat phases of the colloidal hard discs, which increases the free volume and the entropy of the depletants. Such striking quadruple coexistences were identified with the help of the ideas put forward by Bates and Frenkel nearly two decades ago.

The isostructural phase coexistences of ordered phase states of the discs (N$_1$-N$_2$ and C$_1$-C$_2$) are driven by short-ranged attractions (small $q$ and low depletant concentrations). This may be envisaged as an effective `sticky hard platelet' interaction, which is supported by the presence of columnar phase equilibria (C$_1$-C$_2$, I-C$_1$-C$_2$, N-C$_1$-C$_2$, I-N-C$_1$-C$_2$) for relatively small depletants. On the other hand, the isotropic isostructural coexistence is driven by a large depletion zone that sufficiently smoothens the interacting platelet volume. Hence, the relative size of the depletant modifies the coexistence landscape, enhancing isostructural coexistences between partially crystalline phases (C$_1$-C$_2$ and N$_1$-N$_2$) for small depletant sizes while promoting isotropic fluid-fluid (I$_1$-I$_2$)) coexistence for large enough depletants.

\section*{Acknowledgments}
We thank prof. Bert de With, dr. Jos Laven, dr. Gert Jan Vroege,  dr. Heiner Friedrich, and dr. Catarina de Carvalho Esteves for encouraging discussions. We acknowledge prof. Benjamin Widom for a highly insightful discussion about Gibbs' phase rule. AGG thanks Jasper Landman for his input on the graphical abstract. We also acknowledge NWO, DSM, and SymoChem for funding  NWO-TA project $731.015.205$.

\bibliographystyle{tfo}
\bibliography{biblio}

\section{Appendices}
\subsection{Thermodynamics of pure platelet suspensions}\label{ApPlatesExp}
Various thermodynamic properties of pure platelet suspensions have been studied in detail previously\cite{wensink_phase_2009,van_anders_understanding_2014}, and we solely report here the key ingredients required to calculate the final phase diagrams of model colloidal platelet-polymer mixtures. Entropy-driven phase transitions\cite{dijkstra_entropy-driven_2014} as considered here depend on the excluded volume between two colloidal particles. This excluded volume is defined as the volume inaccessible to a second particle in the system as a consequence of the presence of a first  particle\cite{jones_soft_2002}. For two colloidal platelets, the excluded volume ($v_\text{exc}^\text{p-p}$) per particle volume ($v_\text{c}$) reads\cite{onsager_effects_1949}:
\begin{align}
\frac{v_\text{exc}^\text{p-p}}{v_c} &= 2 \left(\left| \cos \gamma \right| +\frac{4 E\left(\sin\gamma \right)}{\pi }+1\right)+\frac{8 \Lambda  \sin \gamma}{\pi }+\frac{2 \sin\gamma}{\Lambda } \quad \text{,}
\end{align}
where $\gamma$ defines the relative orientation between two colloidal discs, and $E(x)$ is the complete elliptic integral of the second kind. Considering the symmetry of a platelet, its orientation can be defined via a unit vector ($\hat{u}$) in the axis of symmetry of the cylinder (hence, $\cos\gamma = \hat{u}\cdot\hat{u}'$). 

For the isotropic and nematic phases we consider Onsager-Parsons-Lee theory\cite{wensink_isotropicnematic_2001,wensink_phase_2009,lekkerkerker_multiphase_2015}. The free energy of both the isotropic and nematic phases reads:
\begin{align}
\frac{\widetilde{F}_k}{\phi_\text{c}} = 
    \ln{\widetilde{v}_\text{c}}
        +
    \ln\phi_\text{c}-1
        +
    \sigma_k[f(\hat{u})]
        + 
    \frac{2}{\pi}\frac{\phi_\text{c}}{\Lambda}G_\text{P}(\phi_\text{c})\langle\langle \Theta_\text{Exc}(\hat{u},\hat{u}') \rangle \rangle
        \quad \text{.}
\label{eq:OPLFreeEn}
\end{align}
The first two terms on the right-hand-side of Eq. \ref{eq:OPLFreeEn} correspond to the finite-volume normalization of the energy ($\widetilde{v}_\text{c}$ being the dimensionless thermal volume of a platelet) and the ideal gas contribution to the free energy.

In order to calculate the free energy of a nematic phase, an orientational distribution function (ODF, $f(\hat{u})$) needs to be accounted for. A system in which particle orientations are taken into account can be envisaged as a multi-component system in which each component corresponds to a possible particle orientation\cite{onsager_effects_1949,roij_isotropic_2005}. Hence, the ODF is a measure of the probability of finding a particle with a given orientation $\hat{u}$. The rotational entropy term,  $\sigma_k[f(\hat{u})]$ is defined as\cite{vroege_phase_1992,vroege_theory_1993}:
\begin{align*}
\sigma_k[f(\hat{u})] &= \int {f(\hat{u})\ln[4\pi f(\hat{u})]\text{d}\hat{u}} \quad \text{.}
\end{align*}
The dimensionless ensemble-averaged excluded volume follows from\cite{lekkerkerker_multiphase_2015}:
\begin{align*}
\langle\langle \Theta_k^\text{Exc}(\hat{u},\hat{u}') \rangle \rangle &= \frac{1}{D^3} \int{\int{ f(\hat{u}) f(\hat{u}') v_\text{exc}^\text{p-p}(\hat{u}\cdot\hat{u}')\text{d}\hat{u}  \text{d}\hat{u}'}}  \quad \text{.}
\end{align*}
Finally, effects beyond the second osmotic virial coefficient are accounted for in an approximate manner via the Parsons-Lee scaling factor\cite{parsons_nematic_1979,lee_numerical_1987}:
\begin{align*}
G_\text{P} (\phi_\text{c}) &= \frac{4-3 \phi_\text{c} }{4 (1-\phi_\text{c} )^2} \quad\text{.}
\end{align*}
Formally, at each platelet concentration the free energy of the system must be minimized with respect to the ODF, $f(\hat{u})$. Analytical expressions for the ODF can be obtained for the  isotropic state by considering equiprobability of orientations: $f(\hat{u}) = 1/(4\pi)$. In this case, $\sigma_\text{I}[f(\hat{u})] = 0$, and by applying the so-called isotropic averages ($\langle\langle \sin\gamma \rangle\rangle_\text{I} = \pi/4 $, $\langle\langle E\{\sin\gamma\} \rangle\rangle_\text{I} = \pi^2/8 $, and $ \langle\langle \cos\gamma \rangle\rangle_\text{I} = 1/2 $), the free energy of an isotropic ensemble of discs can be written as\cite{wensink_phase_2009,lekkerkerker_multiphase_2015}:
\begin{align}
\frac{\widetilde{F}_\text{I}}{\phi_\text{c}} = 
\ln{\widetilde{v}_\text{c}} +
\ln\phi_\text{c}-1 +
\frac{2}{\pi}\frac{\phi_\text{c}}{\Lambda}G_\text{P}(\phi_\text{c})\Theta_\text{I}^\text{Exc}(\Lambda) \quad \text{,}
\label{eq:FreeEnIsoPlate}
\end{align}
with:
\begin{align*}
\sigma_\text{I}(\phi_\text{c},\Lambda) &=  0 \quad\text{,}\ 
\text{and } \\
\Theta_\text{I}^\text{Exc} (\Lambda) & = \frac{\pi ^2}{8} + \left(\frac{3 \pi }{4}+\frac{\pi ^2}{4}\right) \Lambda + \frac{\pi  \Lambda ^2}{2}  \quad \text{.}
\end{align*}
Furthermore, closed expressions for the free energy of the nematic phase can be obtained via a Gaussian approximation\cite{odijk_theory_1986} for $f(\hat{u})$, based upon considering that all relative orientations can be defined as a Gaussian perturbation from the nematic director vector provides a closed-form expression for the free energy\cite{wensink_phase_2009,lekkerkerker_multiphase_2015}. This Gaussian ODF
\begin{align*}
f_\text{N-G}(\theta) &= \frac{\kappa}{4\pi}\exp\left[-\frac{1}{2}\kappa\theta^2\right]     \quad \text{,}
\end{align*}
where $\theta$ is the polar angle between the nematic director and the orientation of the platelet. Minimizing the free energy with respect to the unknown parameter of the Gaussian ODF  ($\kappa$) provides a closed form for the free energy of the nematic phase\cite{tuinier_phase_2007,wensink_phase_2009}.
\begin{align}
\frac{\widetilde{F}_\text{N-G}}{\phi_\text{c}} = 
\ln{\widetilde{v}_\text{c}} +
\ln\phi_\text{c}-1 +
\ln [\sigma_\text{N-G}(\kappa)]-1 +
\frac{2}{\pi}\frac{\phi_\text{c}}{\Lambda}G_\text{P}(\phi_\text{c})\Theta_\text{N-G}^\text{Exc}(\phi_\text{c},\Lambda) \quad \text{,}
\label{eq:NemCanon}
\end{align}
with:
\begin{align*}
\sigma_\text{N-G}(\phi_\text{c},\Lambda) &=  \ln [\kappa(\phi_\text{c},\Lambda)]-1    \quad \text{,} \\
\Theta_\text{N-G}^\text{Exc} (\Lambda) &= \frac{1}{2}\pi ^{3/2}{\kappa(\phi_\text{c} ,\Lambda )}^{-1/2}+2 \pi  \Lambda \quad \text{, and}\\
\kappa(\phi_\text{c},\Lambda) &= \frac{\pi \phi_\text{c}^2 G_\text{P}^2(\phi_\text{c})}{4\Lambda} \quad\text{.}
\end{align*}

For the columnar phase, a modified Lennard-Jones-Devonshire (LJD) cell-theory\cite{lennard-jones_critical_1939,kirkwood_statistical_1941,wood_note_1952,barker_theory_1975} approach provides a closed expression for the free energy (see\cite{wensink_equation_2004,wensink_phase_2009}):
\begin{align}
\begin{split}
\frac{\widetilde{F}_\text{C}}{\phi_\text{c}} &= 
\ln{\widetilde{v}_\text{c}} +
\ln\phi_\text{c} -1 
-2 \ln \left[1-\frac{1}{\text{$\Delta $C}(\phi_\text{c} )}\right] \\
&+2 \ln \left[\frac{3\text{$\Delta $C}(\phi_\text{c} )^2 \phi^*(\phi_\text{c} )}{2 \Lambda  \left(1-\text{$\Delta $C}(\phi_\text{c} )^2 \phi^*(\phi_\text{c} )\right)}\right] \\
&-\ln \left[\frac{1}{3} \left(1-\text{$\Delta $C}(\phi_\text{c} )^2 \phi^*(\phi_\text{c} )\right)\right]-2 \quad \text{,}
\end{split}
\label{eq:FCol}
\end{align}
with :
\begin{align*}
\text{$\Delta $C}(\phi_\text{c}) &= \frac{\sqrt[3]{2} K(\phi_\text{c} )^{2/3}-\sqrt[3]{3} 4 \phi^*(\phi_\text{c} )}{6^{2/3} \sqrt[3]{K(\phi_\text{c} )} \phi^*(\phi_\text{c} )}\quad\text{,} \\
K(\phi_\text{c}) &= \sqrt{3 \phi^*(\phi_\text{c} )^3 (243 \phi^*(\phi_\text{c} )+32)}+27 \phi^*(\phi_\text{c} )^2 \quad\text{,} \\
\phi_d^*(\phi_\text{c} ) &= \frac{\phi_\text{c}}{\phi_\text{c}^\text{max}} \qquad\text{, and }\quad 
\phi_\text{c}^\text{max} = \frac{\pi}{2\sqrt{3}} \approx 0.907 \quad \text{.}
\end{align*}
For convenience, we use dimensionless units for the chemical potentials and osmotic pressures:
\begin{align*}
\widetilde{\mu}_k^\text{o} = \beta{\mu}_k^\text{o} \quad\text{;}\quad
\widetilde{\Pi}_k^\text{o} =  \beta{\Pi}_k^\text{o}v_\text{c} \quad \text{,}
\end{align*}
which follow from $\widetilde{F}_k$ as:
\begin{align}
\widetilde{\mu}_k^\text{o}=\left(\frac{\partial\widetilde{F}_k}{\partial{\phi_\text{c}}}\right)_{T,V} 
\quad\text{;}\quad
 \widetilde{\Pi}_k^\text{o} = {\phi_\text{c}}\widetilde{\mu}_k^\text{o} - \widetilde{F}_k^\text{o} 
\label{eq:CanocRel}
\end{align}
For the isotropic phase, Eqs. \ref{eq:FreeEnIsoPlate} and \ref{eq:CanocRel} provide: 
\begin{align}
\begin{split}
\widetilde{\mu}_\text{I}^\text{o} &= \widetilde{\mu}_\text{I}^\text{*}  +
\ln (\phi_\text{c} ) +\frac{\phi_\text{c}  \left(4 \Lambda ^2+2 [3+\pi ] \Lambda +\pi \right)}{2 (1-\phi_\text{c} )^3 \Lambda } \\
&+\frac{3 \phi_\text{c} ^3 \left(4 \Lambda ^2+2 [3+\pi] \Lambda +\pi \right)}{16 (1-\phi_\text{c})^3 \Lambda }-\frac{9 \phi_\text{c} ^2 \left(4 \Lambda ^2+2 [3+\pi] \Lambda +\pi \right)}{16 (1-\phi_\text{c})^3 \Lambda } \\
\widetilde{\Pi}_\text{I}^\text{o} &=\frac{1}{(1-\phi_\text{c})^3}
\left[
\phi_\text{c}+\phi_\text{c}^2\left(\Lambda-\frac{3}{2}+\frac{\pi(1+2\Lambda)}{4\Lambda}\right) \right] 
\\&
+\frac{ \phi_\text{c}^3}{(1-\phi_\text{c})^3}
\left(\frac{9}{4}-\frac{\Lambda}{2} - \frac{\pi(1+2\Lambda)}{8\Lambda}\right)
\end{split}
\label{eq:Ap1}
\end{align}
In the nematic phase, using the Gaussian approximation, we obtain from Eqs. \ref{eq:NemCanon} and \ref{eq:CanocRel}:
\begin{align}
\begin{split}
\widetilde{\mu}_\text{N-G}^\text{o} &= \widetilde{\mu}_\text{N-G}^\text{*} +
\frac{12-3 \phi_\text{c} -23 \phi_\text{c} ^2+22 \phi_\text{c} ^3 -6 \phi_\text{c} ^4}{(1-\phi_\text{c} )^3 (4-3 \phi_\text{c} )}\\
&+\ln \left[\frac{\pi  (4-3 \phi_\text{c} )^2 \phi_\text{c} ^2}{(\phi_\text{c} -1)^4 \Lambda ^2}\right]+\ln (\phi_\text{c} )-6 \log (2) \\
\widetilde{\Pi}_\text{N-G}^\text{o} &= \frac{12 \phi_\text{c}-19 \phi_\text{c} ^2+17 \phi_\text{c} ^3 -11 \phi_\text{c} ^4+3 \phi_\text{c} ^5 }{(1-\phi_\text{c} )^3 (4-3 \phi_\text{c} )} \quad\text{.}
\end{split}
\label{eq:Ap2}
\end{align}
In view of the complicated expression for the free energy of the columnar phase, we determined the chemical potential and the osmotic pressure in a numerical form\cite{wensink_phase_2009}. All numerical solutions of the analytical expressions were obtained using \textit{Wolfram Mathematica}\cite{WOLF}. 

\subsection{Free volume fraction}\label{sec:ResAlf}
\label{Ap:alfa}
The free volume fraction is related to $W$ via:
\begin{align*}
\alpha = \frac{\langle V_\text{free} \rangle_{\text{o}}}{V} = e^{-\beta W} \quad ,
\end{align*}
where ${\langle V_\text{free} \rangle_{\text{o}}}$ is the free volume for depletants in the undistorted (depletant-free) system, and $V$ is the volume of the system. This work ($W$) is obtained by connecting the limits of inserting a very small depletant and a very big depletant in the system of interest, followed by scaling back to the actual size of the depletant (the core idea of SPT). In the small depletant limit, a second-order Taylor expansion is used, and higher order terms are collected via the work obtained for the big depletant insertion limit. For spherical depletants, a scaling factor ($\lambda$) enables to express this work of insertion as: 
\begin{align*}
W &= \lim_{\lambda\to 1} W (\lambda)\quad\text{,} \\
W (\lambda) &= 
\underbrace{
    \vphantom{ \left(\frac{a^{0.3}}{b}\right) }
    W(0)}_\text{point-depletant} +
\underbrace{
\vphantom{ \left(\frac{a^{0.3}}{b}\right) }
\left.\frac{\partial W}{\partial \lambda}\right\vert_{\lambda = 0} \lambda + 
\frac{1}{2} \left.\frac{\partial^2 W}{\partial \lambda^2}\right\vert_{\lambda = 0} \lambda^2 }_\text{small-depletant} +
\underbrace {
\vphantom{ \left(\frac{a^{0.3}}{b}\right) }
v_\text{d}{\Pi}_k^\text{o}}_\text{big-depletant} \hspace{1pt} \text{.}
\end{align*}
In the small depletant insertion ($\lambda \ll 1$) limit, the free volume depends on the (scaled) depletion zones around the colloidal platelets: 
\begin{align*}
\alpha (\lambda \ll 1) &= 1 - \phi_\text{c} \left(\frac{v_\text{exc}(\lambda)}{v_\text{c}}\right) \quad \text{,}
\end{align*}
so
\begin{align*}
\beta W (\lambda \ll 1) &= - \ln{\left[1 - \phi_\text{c} \left(\frac{v_\text{exc}(\lambda)}{v_\text{c}}\right)\right]} \quad\text{,}
\end{align*}
with a scaled depletion volume derived from Eq. \ref{eq:PHSPlateExc} by scaling the depletant thickness ($\delta \rightarrow \lambda\delta$):
\begin{align*}
\frac{v_\text{exc}(\lambda)}{v_\text{c}} &= 1 + 2 q\lambda + q^2\lambda^2+\frac{\lambda q}{\Lambda } +\frac{\pi  \lambda^2 q^2}{2 \Lambda } + \frac{4 q^3}{3 \Lambda } \quad\text{.}
\end{align*}
As expected,  when considering point depletants ($\lambda = 0$) the free volume is simply the space unoccupied by the colloidal particles:
\begin{align*}
\beta W (\lambda = 0) = - \ln{\left(1 - \phi_\text{c}\right)}  \rightarrow \alpha  (\lambda = 0) = 1 - \phi_\text{c} \quad\text{.}
\end{align*}
The remaining terms in the $\lambda \ll 1$ limit can be arranged into a shape-dependent term, $Q_\text{s}$:
\begin{align*}
Q_\text{s}(\phi_\text{c} ,\Lambda,q )  &= \beta
\lim_{\lambda\to 1} \left[\left(\frac{\partial W}{\partial \lambda}\right)_{\lambda = 0} \lambda + \frac{1}{2} \left(\frac{\partial^2 W}{\partial \lambda^2}\right)_{\lambda = 0} \lambda^2 \right] \quad\text{.}
\end{align*}
For platelets with added spherical depletants we find:
\begin{align*}
Q_\text{s}(\phi_\text{c} ,\Lambda,q ) &= q \left(\frac{1}{\Lambda }+\frac{\pi  q}{2 \Lambda }+q+2\right) y(\phi_\text{c}) \\ &
+ 2 q^2 \left(\frac{1}{4 \Lambda ^2} +\frac{1}{\Lambda }+1\right) y(\phi_\text{c})^2 \quad \text{,}
\end{align*}
with
\begin{align*}
y(\phi_\text{c}) = \frac{\phi_\text{c}}{1-\phi_\text{c}} \quad .
\end{align*}
In the big depletant insertion limit ($\lambda \gg 1$), $W$ is related to the work required to create a cavity of the size of the scaled depletant in the system:
\begin{align*}
\beta W (\lambda \gg 1) &= \beta v_\text{d} \Pi_k^\text{o} = \frac{v_\text{d}}{v_\text{c}}\widetilde{\Pi}_k^\text{o} \quad\text{,}
\end{align*}
where we consider $W$ to depend on the specific colloidal phase state. Hence, the subscript $k$ corresponds again to the an I, N, or C phase.
With both contributions at hand, the final free volume fraction reads:
\begin{align*}
\alpha_k(\phi_\text{c},\Lambda,q) &= 
\underbrace{
        \vphantom{ \left(\frac{a^{0.3}}{b}\right) }
    (1-\phi_\text{c} )}_\text{point-depletant} 
\underbrace{
        \vphantom{ \left(\frac{a^{0.3}}{b}\right) }
    \exp \left[-Q_\text{s}(\phi_\text{c} ,\Lambda,q )\right]}_\text{small depl.}
\underbrace{
        \vphantom{ \left(\frac{a^{0.3}}{b}\right) }
    \exp \left[-\frac{v_\text{d}}{v_\text{c}}\widetilde{\Pi}_k^\text{o}(\phi_\text{c} ,\Lambda ) \right] }_\text{cavity} 
    \quad\text{,}
\end{align*}
as shown in Eq. \ref{eq:alfa}.

\subsection{Phase diagram calculation and characteristic points}\label{diagcalc}
With all the components required to calculate the (dimensionless) grand potential at hand, determination of phase coexistence is straightforward. We define a dimensionless osmotic pressure and chemical:
\begin{align*}
\widetilde{\mu}_k = \beta{\mu}_k \quad \text{;} \quad
\widetilde{\Pi}_k =  \beta{\Pi}_k v_\text{c} \quad \text{,}
\end{align*}
and use the  standard thermodynamic relations to yield the chemical potential of the colloids and the osmotic pressure of the colloid-polymer mixture from the grand-potential $\widetilde{\Omega}_k$:
\begin{align}
\widetilde{\mu}_k=\left(\frac{\partial\widetilde{\Omega}_k}{\partial\phi_\text{c}}\right)_{T,V,N_d^R} \quad\text{;}\quad \widetilde{\Pi}_k = \phi_\text{c}\widetilde{\mu}_k - \widetilde{\Omega}_k  \quad\text{.}
\label{eq:GranCanRel}
\end{align}
Coexistence between different phases is found if
\begin{align*}
\widetilde{\Pi}_{i} = \widetilde{\Pi}_{j}\quad
\text{, and}
\quad
\widetilde{\mu}_{i} = \widetilde{\mu}_{j}  \quad \text{,}
\end{align*}
with the subscripts $\{i,j,k\}$ denoting the possible states of the system: isotropic (I), nematic (N), or columnar (C). Further details for $\widetilde{\Pi}_{i}$ and $\widetilde{\mu}_{i}$ are provided in Appendix \ref{ApPlatesPolExp}. As a consequence of the effective attraction due to the addition of depletants, isostructural I$_1$-I$_2$, N$_1$-N$_2$, and C$_1$-C$_2$ coexistences are possible depending on $q$ and $\Lambda$. An isostructural phase coexistence region is marked by a critical point, which can be calculated from the conditions:
\begin{align*}
 \left(\frac{\partial\widetilde{\mu}_k}{\partial\phi_\text{c}}\right)_{T,V,N_d^R} = \left(\frac{\partial^2\widetilde{\mu}_k}{\partial\phi_\text{c}^2}\right)_{T,V,N_d^R} = 0 \quad \text{.}
\end{align*}
When that is the case, it may be also possible that there are three coexisting phases for which the condition:
\begin{align*}
\widetilde{\Pi}_{i} = \widetilde{\Pi}_{j} = \widetilde{\Pi}_{m} \quad
\text{, and} \quad
\widetilde{\mu}_{i} = \widetilde{\mu}_{j} = \widetilde{\mu}_{m}  \quad 
\end{align*}
holds. Extrapolation to four-phase coexistence is straightforward. 

\subsection{Expressions for various FVT functions}\label{ApPlatesPolExp}
The following relations for the grand-canonical functions are straightforward to infer from Eqs. \ref{eq:Omega} and \ref{eq:GranCanRel}:
\begin{align*}
\begin{split}
\widetilde{\Omega}_k(\phi_\text{c},\Lambda,q,\phi_d^\text{R}) &= 
    \widetilde{F}_k(\phi_\text{c} ,\Lambda )- 
    \frac{v_\text{c}}{v_\text{d}}\alpha_k(\phi_\text{c} ,\Lambda ,q )\widetilde{\Pi}_\text{d}^\text{R}(\phi_d^\text{R}) \quad\text{,}
\\
\widetilde{\mu}_k(\phi_\text{c},\Lambda,q,\phi_d^\text{R}) &= 
    \widetilde{\mu}_k^o(\phi_\text{c} ,\Lambda )- 
    \frac{v_\text{c}}{v_\text{d}}\widetilde{\Pi}_\text{d}^\text{R}(\phi_d^\text{R}) \left(\frac{\partial\alpha_k(\phi_\text{c} ,\Lambda ,q )}{\partial \phi_\text{c}}\right)_{\Lambda,q} \quad\text{,}
\\
\widetilde{\Pi}_k(\phi_\text{c},\Lambda,q,\phi_d^\text{R}) &= 
    \widetilde{\Pi}_k^o(\phi_\text{c} ,\Lambda )- 
    \frac{v_\text{c}}{v_\text{d}}\widetilde{\Pi}_\text{d}^\text{R}(\phi_d^\text{R})
\\ &
\left[\phi_\text{c}\left(\frac{\partial\alpha_k(\phi_\text{c} ,\Lambda ,q )}{\partial \phi_\text{c}}\right)_{\Lambda,q}-\alpha_k(\phi_\text{c},\Lambda,q)\right] \quad\text{.}
\end{split}
\end{align*}
The derivative of the free volume fraction $\alpha_k$ becomes relevant in view of its role in the grand-canonical chemical potential and osmotic pressure of the different colloidal phase sates. It follows from Eq.\ref{eq:alfa} as:
\begin{align*}
\left(\frac{\partial\alpha_k(\phi_\text{c} ,\Lambda ,q )}{\partial \phi_\text{c}}\right)_{\Lambda,q} &=
2 (\phi_\text{c} -1) q^3 (\partial\widetilde{\Pi}_k(\phi_\text{c},\Lambda)/\partial\phi_\text{c})_{\Lambda,q}
\\ &+ 3 \Lambda  \left((\phi_\text{c} -1) (\partial Q_s(\phi_\text{c} ,q,\Lambda )/\partial\phi_\text{c})_{\Lambda,q}-1\right) \text{,}
\end{align*}
with:
\begin{align*}
\left(\frac{\partial Q_\text{s}(\phi_\text{c} ,\Lambda,q )}{\partial\phi_\text{c}}\right)_{\Lambda,q}&= 
\frac{(2 \Lambda +1) q}{(1-\phi_\text{c})^2 \Lambda } \\
&+ \frac{q^2 (\phi_\text{c}  ((-6 \Lambda +\pi -8) \Lambda -2)-\Lambda  (2 \Lambda +\pi ))}{2 (\phi_\text{c} -1)^3 \Lambda ^2} \text{.}
\end{align*}
The derivatives of the canonical expressions are easily obtained from the formulas presented in the previous appendices (Eqs. \ref{eq:Ap1} and \ref{eq:Ap2}).
\appendix

\end{document}